\documentclass[aps,prb,twocolumn,superscriptaddress]{revtex4}

\usepackage{graphicx}	%gestion des figures et des parametres graphiques comme columnwidth, des eps...
\usepackage{amssymb}	%plus de symboles spéciaux
\usepackage{amsfonts}	%gestion des fonts dans les formules
\usepackage{amsmath}	%ajout de possibilités dans les formules, comme \substack par ex
\usepackage{xcolor}		%gestion des couleurs
\usepackage[T1]{fontenc}	%gestion des accents
\usepackage[utf8]{inputenc} %gestion des accents
\usepackage{textcomp} %gestion de \textdegree pour les degres celsius dans la biblio
\usepackage{makecell} %gestion des retours à la ligne dans les cellules des tables
\usepackage{booktabs} %meilleure gestion des tables, donne accès à toprule, midrule etc

\usepackage[english]{babel} 	%gestion propre de la typographie
\usepackage[colorlinks=true,linkcolor=blue,citecolor=blue]{hyperref} %gestion des references et liens internes et externes

\usepackage[caption=false]{subfig}		%gestion des sous-figures, le package caption est incompatible avec revtex
\usepackage{braket}		%ajout des bra et kets dans les formules
\usepackage{bm}		%gestion du bold dans les formules
\usepackage[load-configurations = abbreviations]{siunitx}	%gestion des unités

\begin{document}

\title{Success and Breakdown of the T-Matrix Approximation for Phonon-Disorder Scattering}

\author{S. Th\'ebaud}
\email[E-mail: ]{thebaudsj@ornl.gov}
\affiliation{Materials Science and Technology Division, Oak Ridge National Laboratory, Oak Ridge, Tennessee 37831, USA}
\author{C. A. Polanco}
\affiliation{Materials Science and Technology Division, Oak Ridge National Laboratory, Oak Ridge, Tennessee 37831, USA}
\author{L. Lindsay}
\affiliation{Materials Science and Technology Division, Oak Ridge National Laboratory, Oak Ridge, Tennessee 37831, USA}
\author{T. Berlijn}
\email[E-mail: ]{berlijnt@ornl.gov}
\affiliation{Center for Nanophase Materials Sciences, Oak Ridge National Laboratory, Oak Ridge, Tennessee 37831, USA}
\affiliation{Computational Sciences and Engineering Division, Oak Ridge National Laboratory, Oak Ridge, Tennessee 37831, USA}

\begin{abstract}
We examine the validity of the widely used T-matrix approximation for treating phonon-disorder scattering by implementing an unfolding algorithm that allows simulation of disorder up to tens of millions of atoms. The T-matrix approximation breaks down for low energy flexure phonons that play an important role in thermal transport in two-dimensional materials. Furthermore, insights are developed into the success of the T-matrix approximation in describing maximally mass disordered systems. To achieve this, the phonon unfolding formalism is generalized to describe mass disorder and strongly nonperturbative features of the spectrum are connected to the Boltzmann quasiparticle picture.
\end{abstract}

\maketitle

\section{Introduction} 

Heat management has become an important technological and environmental issue, now critically considered for further electronics miniaturization and for realistic solutions of global energy challenges. As two thirds of the energy consumed worldwide is dissipated as waste heat, \cite{formanEstimatingGlobalWaste2016} the development of more efficient heat management strategies with regards to thermal insulation, transport, and conversion will drive transformative industrial, economic, and environmental impacts. \cite{shinAdvancedMaterialsHighTemperature2019,thekdiIndustrialWasteHeat2015,bellCoolingHeatingGenerating2008,beekmanInorganicCrystalsGlass2017,champierThermoelectricGeneratorsReview2017} However, scientific questions remain regarding thermal transport in materials with defects, which can limit thermal functionalities, but also provide an important tuning mechanism for engineering thermal transport, particularly for thermal insulation and thermoelectric applications. \cite{wagnerSimulationThermoelectricDevices2007,tanHighThermoelectricPerformance2018,muElectronicTransportPhonon2018,kimDenseDislocationArrays2015,kleinkeNewBulkMaterials2010,fuCollectiveGoldstonemodeinducedUltralowLattice2018,cahillLowerLimitThermal1992,nolasTransportPropertiesPolycrystalline2007}

Widely adopted first principles transport methods \cite{broidoIntrinsicLatticeThermal2007,esfarjaniHeatTransportSilicon2011a} have been coupled with combinations of second-order perturbation theory (i.e. Fermi's golden rule \cite{polancoThermalConductivityInN2018,xieBondorderTheoryPhonon2014,tamuraIsotopeScatteringDispersive1983}) and Green's function methods (i.e. the T-matrix approximation \cite{mingoClusterScatteringEffects2010a,katchoEffectNitrogenVacancy2014}) to evaluate the influence of isotopes, \cite{lindsayInitioThermalTransport2013,lindsayPhononThermalTransport2014,lindsayEnhancedThermalConductivity2011} vacancies, \cite{xieBondorderTheoryPhonon2014,polancoDefectlimitedThermalConductivity2020,polancoInitioPhononPoint2018} atomic substitutions, \cite{polancoThermalConductivityInN2018,dongreResonantPhononScattering2018} and mass and force constant disorder \cite{arrigoniFirstprinciplesQuantitativePrediction2018a,kunduRoleLightHeavy2011,gargRoleDisorderAnharmonicity2011a,leeLatticeThermalConductivity2014,tianPhononConductionPbSe2012,liThermalConductivityBulk2012,schradeRoleGrainBoundary2017a,shiomiThermalConductivityHalfHeusler2011,maIntrinsicThermalConductivities2016} on thermal transport in nanoscale, bulk, and alloy systems. 
The T-matrix approximation, in particular, is computationally accessible, even from first principles density functional theory methods, includes resonant scattering effects, and is not limited by linewidth resolution, which can be crucial for calculating thermal transport contributions from low energy, long mean free path phonons. However, success of perturbative methods and the T-matrix approximation in determining thermal conductivities of maximally-disordered alloys \cite{gargRoleDisorderAnharmonicity2011a,liThermalConductivityBulk2012,maIntrinsicThermalConductivities2016} is surprising due to their strong mass variations and defect concentrations well beyond the dilute limit. Furthermore, phonon unfolding methods in alloys with large mass differences have demonstrated strongly altered phonon spectra, including the appearance of heavy and light branches where only acoustic branches exist in the virtual crystal approximation (VCA). \cite{kormannPhononBroadeningHigh2017} Such strong distortions of the vibrational structure are difficult to reconcile with approaches based in the phonon quasiparticle picture.

\begin{figure}
\centering
\includegraphics[width=1.0\columnwidth]{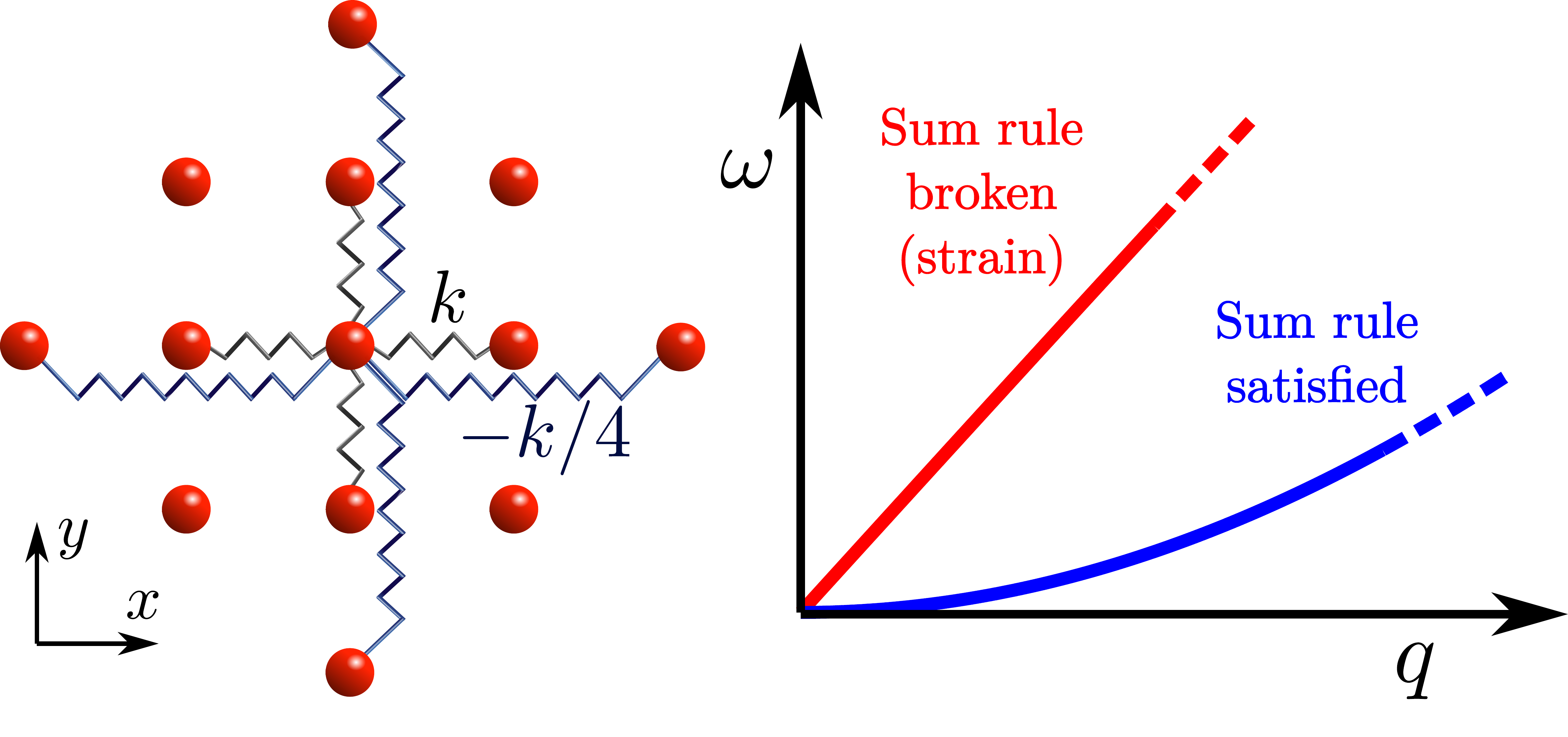}
\caption{Sketch of a 2D out-of-plane spring model that gives rise to a quadratic flexure dispersion (blue curve). Breaking of rotational sum rules (i.e. with residual forces or strain) gives a linear dispersion (red curve).}
\label{fig1}
\end{figure}

An important class of crystals for which the validity of the independent defect scattering approximation may be questionable is two-dimensional (2D) and van der Waals (vdW) layered materials. These systems have received significant attention over the past decade due to their interesting electronic and vibrational properties and potential for device applications. \cite{phamRecentAdvancesDoping2016,radisavljevicSinglelayerMoS2Transistors2011,lembkeBreakdownHighPerformanceMonolayer2012} Monolayer materials feature peculiar out-of-plane modes characterized by quadratic dispersion near the Brillouin zone center $\Gamma$, reflecting the ease with which the atomic sheet can vibrate out of the plane. This can be illustrated by a simple two-dimensional spring-mass model sketched in Fig.~\ref{fig1}, featuring effective out-of-plane springs between the nearest and third-nearest neighbors. The latter negative spring constants are required to satisfy the rotational sum rules (which are equivalent to the absence of residual forces and stress) that give rise to quadratic dispersion. \cite{sarkarBornhuangInvarianceConditions1977} Without these, the sum rules are broken and the flexure branch has linear dispersion as in a strained system. These quadratic out-of-plane modes can carry a large proportion of the heat in 2D and vdW systems, \cite{lindsayFlexuralPhononsThermal2010,liThermalConductivityPhonon2013,seolTwoDimensionalPhononTransport2010} and both T-matrix and phonon unfolding studies of defect scattering in graphene suggest that they are sensitive to disorder, particularly vacancies. \cite{bouzerarDrasticEffectsVacancies2020,polancoInitioPhononPoint2018}However, evaluation of the validity of the T-matrix approximation for phonon-vacancy scattering of the quadratic modes has not been carried out. Such an evaluation is important as prominent monolayer compounds such as MoS$_2$ can have large vacancy concentrations (beyond the dilute limit); a few percents are not atypical depending on synthesis methods, and up to $20 \%$ is possible if engineered. \cite{hongExploringAtomicDefects2015,komsaPointExtendedDefects2013,mahjouri-samaniTailoringVacanciesFar2016} More generally speaking, developing insights into the range of validity (strength of perturbations and defect concentrations) for prominent T-matrix methods is critically important for building confidence in defect-limited first principles transport predictions.

In this article, the validity of the T-matrix approximation is evaluated in generic two and three-dimensional models with vacancies and strong mass disorder. To this end, the phonon spectrum of large disordered supercells of tens of millions of atoms is unfolded using the non-perturbative Chebyshev polynomials Green's function method.\cite{ferreiraCriticalDelocalizationChiral2015} We demonstrate that low-energy out-of-plane vibrations in 2D monolayers are especially sensitive to multiple-impurity scattering effects. We find that the usual unfolding procedure to calculate the spectral function must be generalized to treat mass disorder, and that strong mass variance gives rise to disorder-induced dispersive branches. We connect these results to the Boltzmann transport picture to understand the success of perturbative and T-matrix methods in predicting the thermal conductivity of bulk alloys despite strong distortions of their phonon spectra. 

This manuscript is organized as follows: the T-matrix approximation and phonon unfolding method are briefly explained in section~\ref{methods}. Section~\ref{2d_vacancies} focuses on phonon-defect scattering in atomic monolayers featuring vacancies. In section~\ref{mass_disorder} is discussed the case of mass disordered alloys. We summarize our study in section~\ref{summary}.

\section{Methods} 
\label{methods}

We focus here on three approaches for calculating phonon-defect scattering rates, Fermi's golden rule (FGR), the T-matrix approximation (TMA), and the supercell phonon unfolding method (SPU). For other technical approaches we refer to Ref.~\onlinecite{muUnfoldingComplexityPhonon2020,ghoshPhononsRandomAlloys2002,shiomiThermalConductivityHalfHeusler2011,mcgaugheyQuantitativeValidationBoltzmann2004}.

The T-matrix method approximates the self-energy of the Green's function by a subset of diagrams corresponding to successive phonon scatterings with a single impurity. \cite{aiyerPairEffectsSelfConsistent1969} As such, it is most appropriate in the limit of low defect concentrations and is expected to break down when multiple-impurity scatterings become important. For TMA calculations presented here, the scattering rate of the Bloch mode with wavevector $q$ is given by \cite{mingoClusterScatteringEffects2010a,katchoEffectNitrogenVacancy2014,economouGreenFunctionsQuantum1979}
\begin{equation}
\label{lifetime_tmatrix}
\frac{1}{\tau_q} = -\frac{x}{\omega^0_q}  \, \text{Im} \left( \sum_{r r'} e^{-i q \cdot r} T_{r r'}(\omega^0_q) e^{i q \cdot r'} \right) ,
\end{equation}
where $r$ runs through the atoms, $x$ is the impurity concentration, $\omega^0_q$ is the frequency of mode $q$ in the clean system, and  $T$ is the T-matrix: $T (\omega) = P(\omega) [1 - G^0(\omega) P(\omega)]^{-1}$. $P$ is the perturbation of a single impurity: $P_{r r'}(\omega) = \Delta \phi_{r r'}/m^0 - \omega^2 \delta_{r r'} \Delta m_r/m^0$, with $\Delta \phi_{r r'}$ the change in the interatomic force constant (IFC) between $r$ and $r'$, and $\Delta m_r = m_r - m^0$ is the mass variation of $r$. $G^0(\omega) = [(\omega + i \eta)^2 - D]^{-1}$ is the Green's function of the clean system, with $D_{r r'} = \phi_{r r'}/m^0$ the dynamical matrix and $\eta$ an infinitesimally small positive imaginary part. The direction indices have been dropped as different directions are decoupled in our models. Due to the local nature of the perturbation, the dimensions of $P$ and $T$ are relatively small and matrix inversions and multiplications are straightforward. To obtain the scattering rate from FGR, the T-matrix in Eq.~\eqref{lifetime_tmatrix} is replaced with the second-order perturbation correction: $P(\omega) + P(\omega) G^0(\omega) P(\omega)$.

In recent years, headway has been made in the study of vibrations in strongly disordered materials via the supercell phonon unfolding method.\cite{baroniPhononDispersionsMathrmGa1990a,allenRecoveringHiddenBloch2013,boykinBrillouinZoneUnfolding2014,delaireHeavyimpurityResonanceHybridization2015,overyDesignCrystallikeAperiodic2016,overyPhononBroadeningSupercell2017,ikedaModeDecompositionBased2017,kormannPhononBroadeningHigh2017,muUnfoldingComplexityPhonon2020,polancoDefectlimitedThermalConductivity2020,bouzerarDrasticEffectsVacancies2020} In this non-perturbative numerical technique, the phonon spectral function is calculated for a disordered supercell according to:
\begin{equation}
\label{simple_unfolding}
A(q,\omega) =  \sum_J  \left| \sum_{r} \frac{e^{-i q \cdot  r}}{\sqrt{N}} V_{r,QJ} \right|^2  \delta(\omega - \omega_{QJ}) ,
\end{equation}
where $r$ runs through the $N$ atoms in the supercell, $Q$ is the wavevector in the supercell first Brillouin zone that is equivalent to $q$ in the supercell reciprocal lattice, $\omega_{QJ}$ is the supercell eigenmode frequency of band $J$ at wavevector $Q$, and $V_{r,QJ}$ is the corresponding eigenvector. For the case of force constant disorder the phonon lifetimes can be extracted from the full-width at half-maximum of the spectral peaks given by Eq.~\eqref{simple_unfolding} (see appendix~\ref{appendix_spectralfunction}). For the case of mass disorder, this correspondence breaks down as will be discussed below. With increasing supercell size, the unfolded spectral function converges to the exact spectral function of the fully disordered system. In practice, the spectral function is often computed through an exact diagonalization of the supercell dynamical matrix. However, the unfavorable $N^3$ scaling of this approach constrains the supercell size to around $\num{d4}$ atoms or less. The resulting finite-size effects in turn limit the energy resolution attainable. This is especially problematic for phonons because the linewidths of the acoustic modes become vanishingly small as $q \rightarrow \Gamma$ and these often contribute significantly to the thermal conductivity. To circumvent this limitation, the spectral function is evaluated here via the Chebyshev polynomial Green's function method. \cite{ferreiraCriticalDelocalizationChiral2015,mayouRealSpaceApproachElectronic1995,weisseKernelPolynomialMethod2006} In this approach, the Green's function is expanded on the Chebyshev polynomial basis (see appendix~\ref{appendix_CPGF}), yielding a linear scaling with respect to $N$. In this way, we are able to reach very large supercell sizes of tens of millions of atoms thus enabling investigation of the low-lying acoustic modes. For such system sizes, averaging of multiple disorder configurations and going beyond $Q=0$ is not necessary.

\begin{figure}
\centering
\includegraphics[width=1.0\columnwidth]{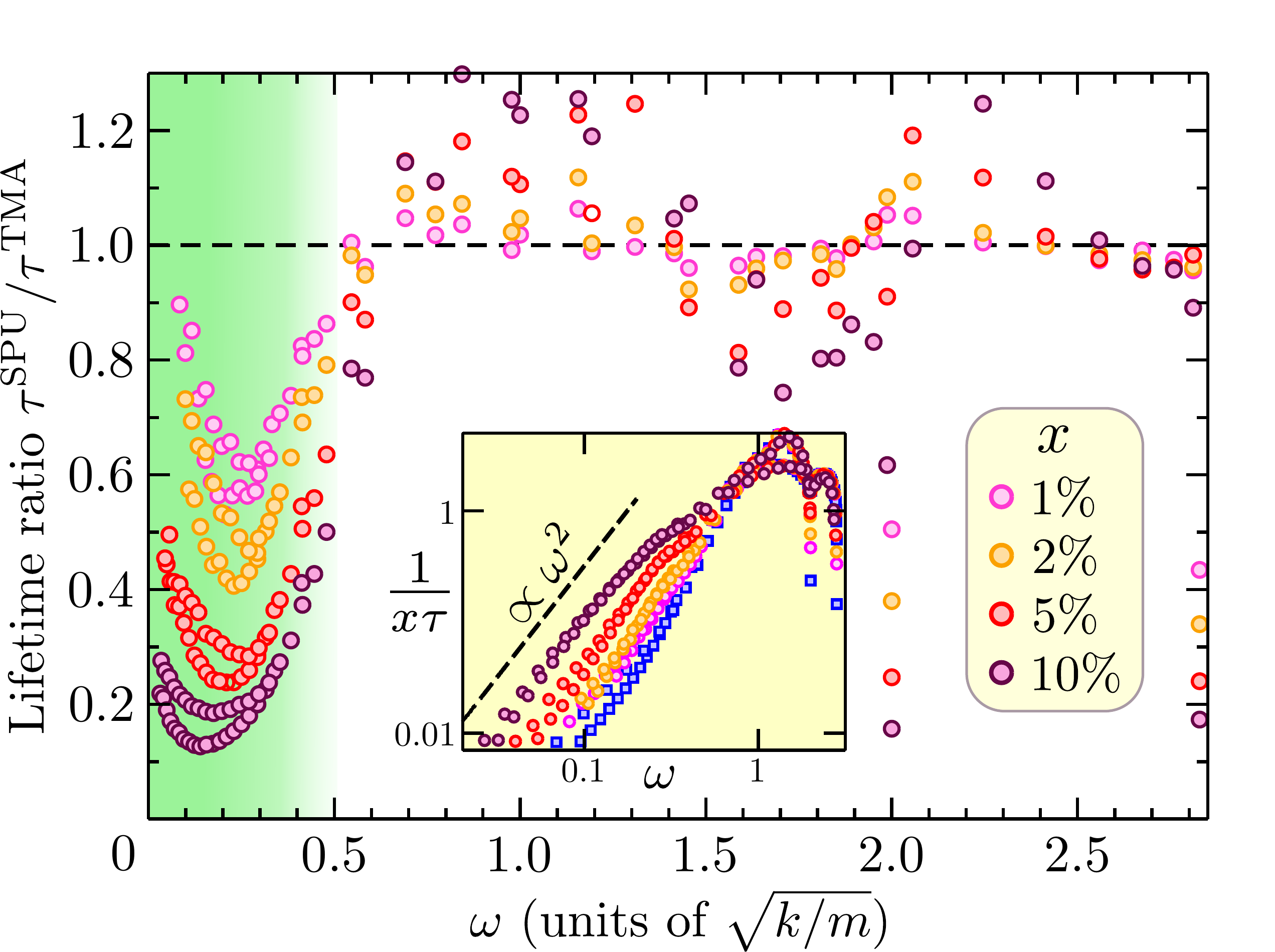}
\caption{(color online) Ratio of phonon lifetimes from supercell phonon unfolding (SPU) and the T-matrix approximation (TMA) in the 2D out-of-plane model (Fig.~\ref{fig1}) for varying vacancy concentrations $x$. Ratios are plotted as a function of the clean mode frequencies. Inset: Scattering rates divided by $x$ as a function of frequency (blue squares correspond to the TMA and circles to the SPU).}
\label{srates_vac}
\end{figure}

\section{Vacancies in 2D systems} 
\label{2d_vacancies}

We consider vacancies in the 2D spring-mass model (Fig.~\ref{fig1}) that exhibits a generic quadratic dispersion. It features atoms of mass $m$ on a square lattice, nearest-neighbor springs of stiffness $k$ and third-nearest neighbor springs of stiffness $-k/4$. All angular frequency and inverse lifetime values are presented in units of $\sqrt{k/m}$. A concentration $x$ of vacancies is introduced by removing the springs between the missing atom and its neighbors. Springs around the vacancies are modified in order to satisfy the rotational sum rules for the IFCs. (see appendix~\ref{appendix_sumrules})

Fig.~\ref{srates_vac} gives the ratio $\tau^{\text{SPU}}_q / \tau^{\text{TMA}}_q$, where $\tau^{\text{TMA}}_q$ and $\tau^{\text{SPU}}_q$ are the disorder-induced phonon lifetimes calculated by the TMA and SPU, respectively, for vacancy concentrations $x=1\%$, $2\%$, $5\%$, and $10\%$. The inset shows the inverse lifetimes divided by $x$. Both methods give a quadratic power law at low frequency, as expected from FGR. \cite{ratsifaritanaScatteringPhononsVacancies1987} The T-matrix predicts the lifetime at angular frequencies above $0.5$ reasonably well, apart from the discrepancies at $2$ due to a Van Hove singularity and at $\approx 2.8$ due to the vanishing of the density of states. However, the T-matrix fails at lower frequencies, especially in the $0.15-0.4$ range. This region, marked by a green shading in Fig.~\ref{srates_vac}, corresponds to the bottom $15\%$ of the out-of-plane spectrum, which typically represents a significant part of the total thermal conductivity in 2D materials (for instance, see appendix~\ref{appendix_SiGe_MoS2} for the case of MoS$_2$). More precisely, in this highly-relevant region, SPU lifetimes can be as low as $55 \%$ of the TMA values for $x = 1\%$, and $40 \%$ for $x = 2\%$. The ratio can reach $25 \%$ for $x = 5\%$ and $15 \%$ for $x = 10\%$. This suggests that the T-matrix approximation breaks down in 2D materials with high vacancy concentrations. Vacancy concentrations larger than $1 \%$ are not uncommon in transition metal dichalcogenides, for instance, which may result in a significant overprediction of defect-limited thermal transport from the TMA. In such cases, accurately predicting the thermal conductivity requires consideration of multiple-impurity scatterings, relying for instance on the SPU. 

\section{Mass disorder} 
\label{mass_disorder}

Mass disorder from isotopes or atomic substitutions presents an important phonon thermal resistance in materials, particularly in alloys such as (Si,Ge) or Mg$_2$(Si,Sn). Mass perturbations are diagonal in real space and proportional to $\omega^2$, which implies that perturbative methods and the TMA should become exact at low energy. \cite{katchoEffectNitrogenVacancy2014,mondalLocalizationPhononsMassdisordered2017} The questions remain: what is the range of validity for these approaches, and does the low frequency behavior explain the success of these techniques in computing thermal conductivities of strongly disordered alloys?

Regarding the SPU method, the phonon unfolding formula~\eqref{simple_unfolding} used throughout the literature\cite{allenRecoveringHiddenBloch2013,ikedaModeDecompositionBased2017,kormannPhononBroadeningHigh2017,delaireHeavyimpurityResonanceHybridization2015,baroniPhononDispersionsMathrmGa1990a,boykinBrillouinZoneUnfolding2014,overyPhononBroadeningSupercell2017} does not probe the spectrum (and thus the lifetime) of a Bloch vibrational mode in the presence of mass disorder. Because the eigenmodes are defined using mass-rescaled atomic displacements, Eq.~\eqref{simple_unfolding} corresponds to a mode in which the true atomic displacements are proportional to $\frac{e^{iq \cdot r}}{\sqrt{N}} \frac{1}{\sqrt{m_r}}$, (see appendix~\ref{appendix_spectralfunction}) with $m_r$ being disordered. The atomic displacements corresponding to actual Bloch modes with ordered masses should be used instead. In the present context, it is helpful to look to the Green-Kubo formula, which is exact beyond the quasiparticle picture in the linear regime without phonon-phonon interactions. Neglecting vertex corrections, the thermal conductivity $\kappa$ can be expressed (see appendix~\ref{appendix_kubo}) as $\kappa\approx \sum_q \frac{\pi \hbar (\omega^0_q v^0_q)^2}{ V T} \int_0^\infty d\hbar \omega  \left( -\frac{\partial f_B}{\partial\hbar\omega} \right) \tilde{A}(q,\omega)^2$ where $T$ is the temperature, $V$ the system size, $f_B$ the Bose-Einstein distribution, $\omega_q^0$ and $v_q^0$ the Bloch mode frequencies and velocities along direction $x$ and
\begin{equation}
\label{unfolding_mass_disorder}
\tilde{A}(q,\omega)=\sum_J  \left| \sum_{r} \frac{e^{-i q \cdot  r}}{\sqrt{N}} \sqrt{\frac{m^0}{m_r}} V_{r,QJ} \right|^2  \delta(\omega - \omega_{QJ}) , 
\end{equation}
is the generalized phonon unfolding formula for the spectral function of mass disordered systems that captures the lifetime of the Bloch mode $q$ (see appendix~\ref{appendix_spectralfunction}). $m^0$ is the mass in the ordered reference system.

\begin{figure}[b]
\centering
\includegraphics[width=1.0\columnwidth]{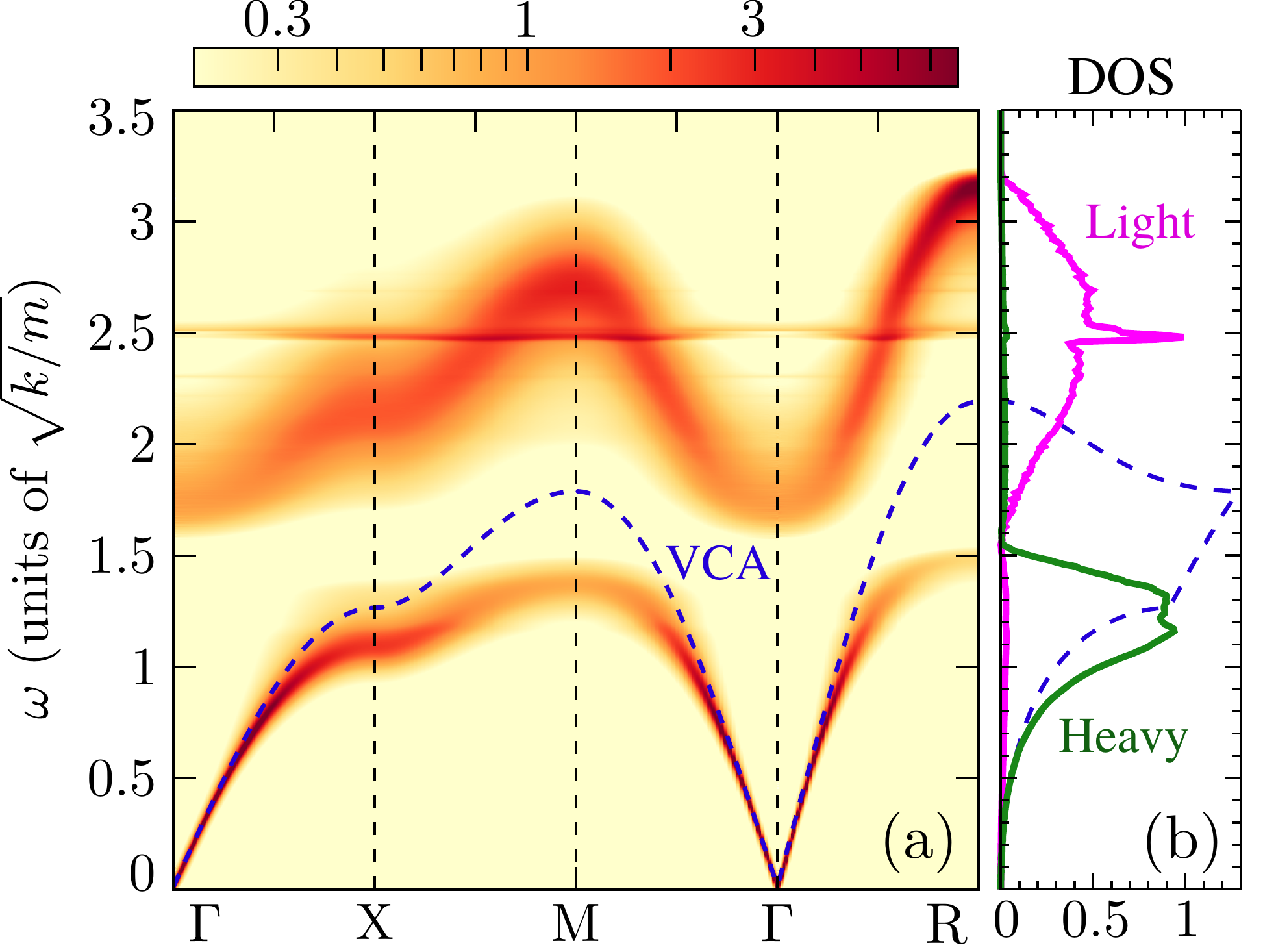}
\caption{(colors online) (a) Phonon spectral function obtained from the generalized unfolding formula for the 3D mass-disordered model with $50\%$ of $m' =4 m$ defects. (b) Phonon DOS projected onto light and heavy atoms. In both panels, the VCA is shown as a dashed blue line.}
\label{spectral_functions}
\end{figure}

Here we assess the effects of mass disorder in bulk materials by calculating the spectral function of a three-dimensional (3D) spring-mass model on a cubic lattice featuring nearest-neighbor springs $k$ and clean atomic masses $m$, with pure mass defects (no IFC perturbations). The mass differences in thermally-relevant alloys can reach a factor four, as in Mg$_2$(Si,Sn) or (In,Al)N alloys. Therefore, we introduce impurities with mass $m' = 4 m$ as a reasonable upper limit to the strength of mass fluctuations and focus on maximally-disordered alloys (half the atoms are substituted). For the reference system we use the VCA: $m^0=\frac{m+m'}{2}$.

Fig.~\ref{spectral_functions}~(a) gives the spectral function $\tilde{A}(q,\omega)$ obtained from the generalized phonon unfolding formula, Eq.~\eqref{unfolding_mass_disorder}. An artificial Lorentizan broadening of $0.003$ has been kept for visualization purposes. Strikingly, the single branch of the clean system separates into an acoustic-like branch at low-energy and an optic-like branch at high energy. Despite these strong disruptions of the vibrational structure, the VCA predicts the phonon dispersion correctly for the acoustic-like branch below $\omega = 0.5$ due to the decrease of the mass perturbation as $\omega$ approaches zero. Fig.~\ref{spectral_functions}~(b) gives the calculated projected density of states (pDOS) for the light and heavy atoms via exact diagonalization of the dynamical matrix of 100 supercells containing 400 atoms on average. The acoustic-like branch is governed primarily by the heavy atom vibrations, while the optic-like branch arises from vibrations of the light atoms against the matrix of heavy atoms (this is qualitatively consistent with calculated vibrational structures of certain binary alloys \cite{kormannPhononBroadeningHigh2017}). Instead of a simple broadening of the VCA branch, two distinct branches are present with equally strong quasiparticle peaks. As these starkly non-perturbative features of the vibrational spectrum demonstrate, the phonon quasiparticle picture that the Boltzmann transport formalism is built upon breaks down, calling into question the validity of previous transport calculations in strongly disordered alloys.

\begin{figure}
\centering
\includegraphics[width=1.0\columnwidth]{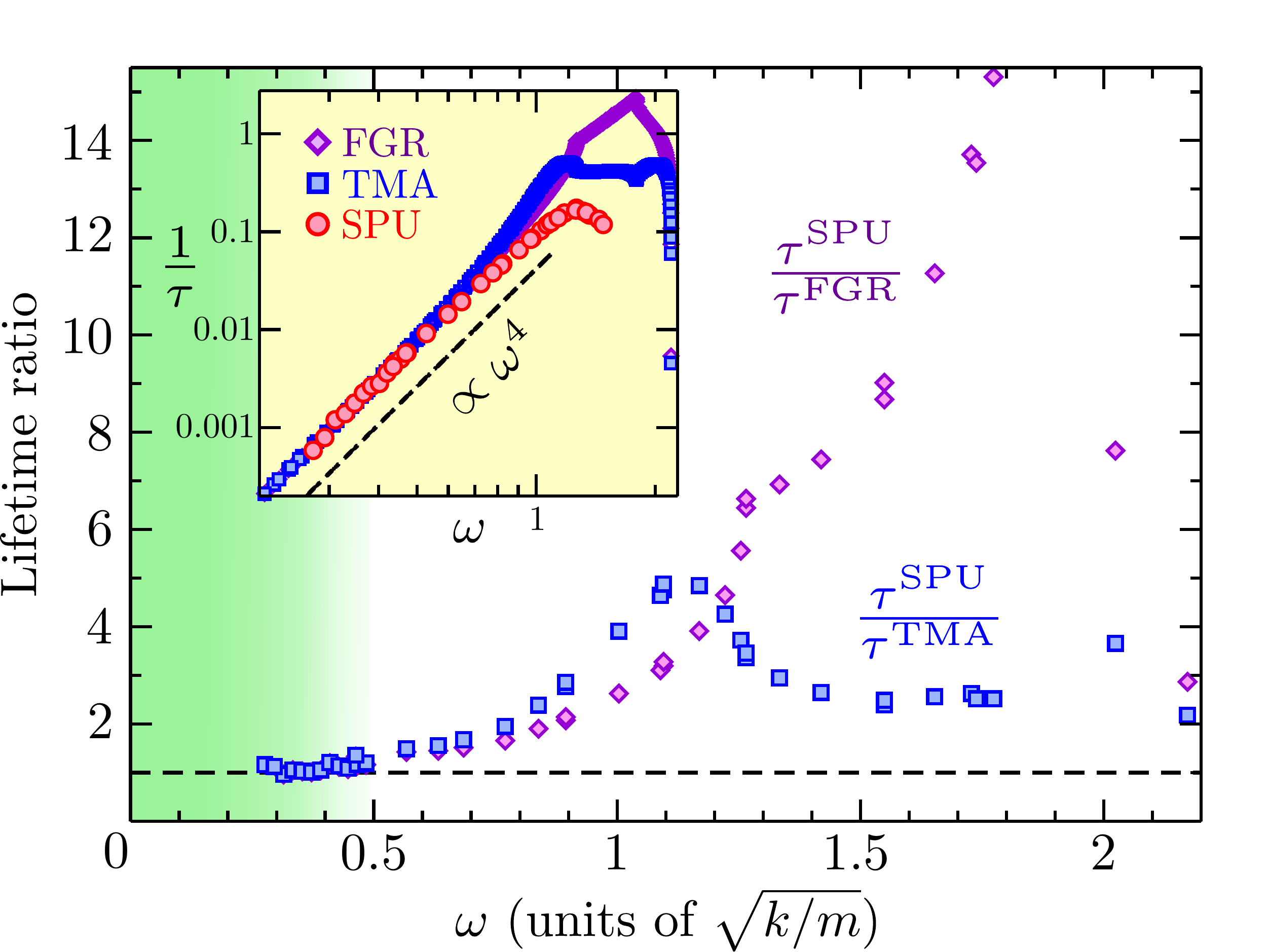}
\caption{(colors online) Phonon lifetimes obtained from supercell phonon unfolding (SPU) relative to those obtained from the T-matrix approximation (TMA) and Fermi's golden rule (FGR) in the 3D mass-disordered model with $50\%$ of $m'=4 m$ defects. The ratios are plotted as a function of the VCA mode frequencies. Inset: scattering rates as a function of frequency.}
\label{srates_3D_mass}
\end{figure}

We assume for simplicity that the spectral function near the resonances can be reasonably approximated by separate Lorentzians: $\tilde{A}(q,\omega) = \sum_{l} F_{ql} \frac{1}{\pi} \frac{1/2 \tau_{ql}}{(\omega - \omega_{ql})^2 + (1/2 \tau_{ql})^2}$, where $l=1,2$ designates the heavy and light branch, respectively, and $F_{ql}$ are quasiparticle weights as given by the spectral function. Using the approximations necessary for derivation of the kinetic equation from the Green-Kubo formula, \cite{elliottTheoryPropertiesRandomly1974} the thermal conductivity can be further simplified:
\begin{equation}
\kappa \approx \sum_{ql} \frac{\hbar}{V T} \left( -\frac{\partial f_B}{\partial \omega} \bigg|_{\omega_{ql}} \right) (F_{ql} \omega^0_q v^0_q)^2 \tau_{ql},  
\end{equation}
This is similar to the kinetic particle description of phonon conductivity for a system with one acoustic branch and one optic branch. Indeed, this equation connects the SPU formalism to the Boltzmann transport theory in the case where strong disorder generates multiple branches. It allows the different peaks to be interpreted as quasiparticle excitations and their broadenings $1/\tau_{ql}$ as inverse lifetimes. Remarkably, the heat flux is given by the products of the VCA frequencies $\omega^0_q$ and velocities $v^0_q$ renormalized with the quasiparticle weights $F_{ql}$, not by the quasiparticle frequencies $\omega_{ql}$ and velocities $\partial \omega_{ql}/\partial q$. 

The broadening of the acoustic-like branch at low energy is well below $0.01$, much smaller than in the rest of the spectrum, including the whole optic-like branch whose broadening is of order $1$. We extracted the inverse phonon lifetimes from the width at half-maximum of the acoustic-like spectral peaks in the high-symmetry segments $\Gamma-X$, $\Gamma-M$ and $\Gamma-R$ of the Brillouin zone. The ratios $\tau^{\text{SPU}}_q / \tau^{\text{TMA}}_q$ and $\tau^{\text{SPU}}_q / \tau^{\text{FGR}}_q$ are shown in Fig.~\ref{srates_3D_mass}, where $\tau^{\text{FGR}}_q$ is the phonon lifetime calculated by FGR \cite{tamuraIsotopeScatteringDispersive1983} (the perturbation is defined with respect to the VCA). The inverse lifetimes themselves are shown in the inset. At high frequencies, the phonon lifetimes are significantly underestimated by the TMA and FGR compared with the SPU values. However, all techniques give an $\omega^4$ power law at low energy, as expected for point defect scattering. There is good agreement (within $15\%$) between all methods below $\omega = 0.4$, which roughly corresponds to the bottom $20\%$ of the VCA frequency range,  as indicated by the shaded region in Fig.~\ref{srates_3D_mass}. These low frequency modes give a significant contribution to heat transport in disordered alloys. In Si$_{0.5}$Ge$_{0.5}$ for example, we find from first-principles calculations that $90\%$ of the thermal conductivity is derived from the bottom $20\%$ of the frequency range (see appendix~\ref{appendix_SiGe_MoS2}). Although definitive pronouncements on any specific material require more realistic calculations, this strongly suggests that the $\omega^2$ dependence of the mass perturbation is the main explanation for the unexpected success of T-matrix and perturbative approaches for calculating alloy thermal conductivity. As we examined the extreme case of large mass ratio and maximal disorder, it is expected that agreement of the different techniques will only improve with weaker disorder. 

\section{Summary} 
\label{summary}

The validity of the T-matrix approximation has been evaluated by comparing disorder-induced lifetimes in generic mass-spring models using the phonon unfolding Chebyshev polynomials Green's function method. This non-perturbative technique allows the study of disordered supercells containing tens of millions of atoms and fully includes multiple-impurity scatterings. In order to include mass variance, we have generalized the unfolding formalism for the spectral function and connected the strongly distorted vibrational structure to the phonon quasiparticle picture using the Green-Kubo formula as a guide. Second-order perturbation theory and the T-matrix approximation are valid at low frequencies in systems dominated by mass disorder, explaining the success of thermal conductivity predictions in various alloys. However, these methods break down in a similarly relevant frequency range for out-of-plane vibrations in two-dimensional systems containing $1\%$ of vacancies or more. In these cases, the phonon unfolding methodology presented in this work captures the relevant multiple-impurity scattering effects.

\textit{Acknowledgements.} This research was supported by the U.S. Department of Energy, Office of Science, Basic Energy Sciences, Materials Sciences and Engineering Division. We used resources of the Compute and Data Environment for Science (CADES) at the Oak Ridge National Laboratory, which is supported by the Office of Science of the U.S. Department of Energy under Contract No. DE-AC05-00OR22725.

\appendix

\section{Two-dimensional model: rotational sum rules and vacancies}
\label{appendix_sumrules}

In two-dimensional materials, in-plane ($xy$) and out-of-plane ($z$) vibrational modes have different characteristic behaviors. The former have a linear dispersion around $\Gamma$, while the latter exhibit a quadratic dispersion. This quadraticity reflects the greater ease with which a two-dimensional atomic layer can vibrate out of the plane than in the plane, and disappears if strain is applied to the material \cite{boniniAcousticPhononLifetimes2012}. In practice, the interatomic force constants (IFC) must satisfy a set of rotational sum rules called the Born-Huang conditions and the Huang invariance \cite{erikssonHiphivePackageExtraction2019}, that are equivalent to the absence of residual forces and strains in the material \cite{sarkarBornhuangInvarianceConditions1977}. To preserve the quadraticity of the phonon dispersion, the out-of-plane IFCs of our two-dimensional model must therefore obey the following constraints:
\begin{equation}
\label{Born-Huang}
\sum_{r'} \phi_{r r'} (r^\alpha - r'^\alpha) = 0
\end{equation}
and
\begin{equation}
\label{Huang}
\sum_{r r'} \phi_{r r'} (r^\alpha - r'^\alpha) (r^{\alpha'} - r'^{\alpha'}) = 0
\end{equation}
where $\alpha, \alpha' = x, y$, $\phi_{r r'}$ is the out-of-plane IFC between atoms $r$ and $r'$, and $r^\alpha$ is component $\alpha$ of the position of atom $r$. We consider a spring-mass model on a square lattice, with atoms of mass $m$ and springs of stiffness $k$ between the first nearest-neighbors. To satisfy condition~\eqref{Huang}, it is necessary to introduce a third nearest neighbor interaction with a spring stiffness $-k/4$. Note also that the IFCs satisfy by construction the translational sum rule $\sum_{r'} \phi_{r r'} = 0$ ensuring the existence of zero-energy modes at $\Gamma$.

When a vacancy is introduced in the 2D model, equations~\eqref{Born-Huang} and~\eqref{Huang} must still be satisfied.  Therefore, we add two modifications of the IFCs: the third-neighbor springs across the vacancies are cut and the spring constants between the four vacancy nearest-neighbors and their first neighbor in the direction opposite the vacancy are reduced by $k/2$. The rotational sum-rules being satisfied, the dispersion relation calculated through the Chebyshev polynomials Green's function method (CPGF) remains quadratic even at high vacancy concentration.

\section{Phonon unfolding and the spectral function}
\label{appendix_spectralfunction}

In a clean, ordered system containing $N_\text{sc} N$ unit cells of a crystal with Born-von Karman periodic boundary conditions, each unit cell contains several atoms $i$ of mass $m^0_{i}$, and $r$ denotes the sites of the Bravais lattice. The displacement of atom $i$ in cell $r$ along direction $\alpha$ from its equilibrium position is given by the quantum operator $\hat{u}^\alpha_{ir}$, and its momentum is given by $\hat{p}^\alpha_{ir}$. $(\phi^0)^{\alpha {\alpha'}}_{ir,i'r'}$ denotes the IFC between atom $i$ in cell $r$ along $\alpha$ and atom $i'$ in cell $r'$ along ${\alpha'}$. The Hamiltonian is
\begin{align}
H^0 & = \sum_{ir \alpha} \frac{(\hat{p}^\alpha_{ir})^2}{2 m^0_{i}} + \frac{1}{2} \sum_{ir \alpha i'r' \alpha'} \hat{u}^\alpha_{ir} (\phi^0)^{\alpha {\alpha'}}_{ir,i'r'} \hat{u}^{\alpha'}_{i'r'} \\
& =  \frac{1}{2} \sum_{ir \alpha} (\hat{P}^{0 \,\alpha}_{ir})^2 + \frac{1}{2} \sum_{ir \alpha i'r' \alpha'} \hat{U}^{0 \, \alpha}_{ir} (D^0)^{\alpha {\alpha'}}_{ir,i'r'} \hat{U}^{0 \, {\alpha'}}_{i'r'} \nonumber
\end{align}
with the reduced coordinates $\hat{U}^{0 \, \alpha}_{ir} = \sqrt{m^0_{i}} \hat{u}^\alpha_{ir}$ and $\hat{P}^{0 \,\alpha}_{ir} = \hat{p}^\alpha_{ir}/\sqrt{m^0_{i}}$ and the clean dynamical matrix $(D^0)^{\alpha {\alpha'}}_{ir,i'r'} = (\phi^0)^{\alpha {\alpha'}}_{ir,i'r'}/\sqrt{m^0_{i} m^0_{i'}}$. Going to the Fourier domain and considering a wavevector $q$ in the first Brillouin zone (BZ) of the lattice yields a $q$-dependent dynamical matrix, which can be diagonalized:
\begin{align}
(D^0_q)^{\alpha {\alpha'}}_{ii'} & = \sum_{r'} \frac{(\phi^0)^{\alpha {\alpha'}}_{i,i' r'}}{\sqrt{m^0_i m^0_{i'}}} e^{i q r'} \nonumber \\
& = \sum_j e^\alpha_{i,qj} (\omega^0_{qj})^2 (e^{\alpha'}_{i',qj})^* 
\end{align}
with $j$ the branch index and $e^\alpha_{i,qj}$ the polarisation vectors, i.e. the normalized eigenvectors of $D^0_q$ with $(\omega^0_{qj})^2$ the corresponding eigenvalues. The Hamiltonian is diagonal in terms of the Bloch displacement and momentum eigenoperators:
\begin{align}
\hat{U}^0_{qj} & = \sum_{i r \alpha} (E^\alpha_{ir,qj})^* \, \hat{U}^{0 \, \alpha}_{ir} \nonumber \\ & =  \frac{1}{\sqrt{N_\text{sc} N}} \sum_{i r \alpha} e^{-i q r}  (e^\alpha_{i,qj})^* \; \hat{U}^{0 \,\alpha}_{ir}
\end{align}
with the momentum eigenoperators defined similarly. $E^\alpha_{ir,qj} = e^{i q r}  e^\alpha_{i,qj} / \sqrt{N_\text{sc} N}$ is a normalized eigenvector of the real-space dynamical matrix $(D^0)^{\alpha {\alpha'}}_{ir,i'r'}$.

Then, let us divide the system in $N_\text{sc}$ supercells (SC) with positions $R$ containing $N$ units each. We introduce mass and IFC disorder ($m^0 \rightarrow m$, $\phi^0 \rightarrow \phi$, $D^0 \rightarrow D$) identically in every SC, so that the system is unchanged by translations from one supercell to another. The Hamiltonian is now
\begin{align}
H & = \sum_{IR} \frac{(\hat{p}^\alpha_{ir})^2}{2 m_{I}} + \frac{1}{2} \sum_{IRI'R'} \hat{u}^\alpha_{IR} \phi^{\alpha {\alpha'}}_{IR,I'R'} \hat{u}^{\alpha'}_{I'R'} \nonumber \\ 
& =  \frac{1}{2} \sum_{IR} (\hat{P}^{\alpha}_{IR})^2 + \frac{1}{2} \sum_{IRI'R'} \hat{U}^{\alpha}_{IR} D^{\alpha {\alpha'}}_{IR,I'R'} \hat{U}^{{\alpha'}}_{I'R'}
\end{align}
with the new reduced coordinates $\hat{U}^{\alpha}_{IR} = \sqrt{m_{I}} \hat{u}^\alpha_{IR}$ and $\hat{P}^{\alpha}_{IR} = \hat{p}^\alpha_{IR}/\sqrt{m_{I}}$, and $I$ running over the atoms in a supercell. The Fourier transformed dynamical matrix can again be defined as a function of $Q$, a wavevector in the reduced BZ of the new Bravais lattice of the supercells:
\begin{align}
\label{DM_supercell}
(D_Q)^{\alpha {\alpha'}}_{II'} & = \sum_{R'} \frac{\phi^{\alpha {\alpha'}}_{I,I' R'}}{\sqrt{m_I m_{I'}}} e^{i Q R'} \nonumber \\ 
& = \sum_J V^\alpha_{I,QJ} \omega_{QJ}^2 (V^{\alpha'}_{I',QJ})^*
\end{align}
with $J$ the new branch index and $ V^\alpha_{I,QJ}$ the normalized eigenvectors of $D_Q$ with $\omega_{QJ}^2$ the corresponding eigenvalues. The displacement eigenoperators are 
\begin{equation}
\hat{U}_{QJ} = \frac{1}{\sqrt{N_\text{sc}}} \sum_{I R \alpha} e^{-i Q R}  (V^\alpha_{I,QJ})^* \; \hat{U}^\alpha_{IR},
\end{equation}
and the Bloch operators can be expressed in terms of the disordered eigenoperators:
\begin{equation}
\hat{U}^0_{qj} = \frac{1}{\sqrt{N}} \sum_{J \, \alpha \, i \, r \in \text{SC}} e^{-i q r}  (e^\alpha_{i,qj})^* \sqrt{\frac{m^0_i}{m_{ir}}} \; V^\alpha_{ir,Q(q) J} \; \hat{U}_{Q(q) J}
\end{equation}
where $r$ runs over the unit cells contained in the central supercell only, and $Q(q)$ is the wavevector in the reduced BZ of the supercell lattice that is equivalent to $q$.

The phonon retarded Green's function of the disordered system can be defined on the eigenbasis from the correlation function of the displacement operators \cite{mahanManyParticlePhysics2000}:
\begin{align}
G_{QJ,Q'J'}(t) & =  \frac{\Theta(t)}{i \hbar}\Braket{0 | \left[ \hat{U}_{QJ}(t), \hat{U}_{Q'J'}^\dagger \right] |0} \nonumber \\ & = - \frac{\Theta(t) \delta_{Q Q'} \delta_{J J'}}{\omega_{QJ}} \sin(\omega_{QJ} t)
\end{align}
with $\Theta$ the Heaviside function and $\ket{0}$ the phonon ground state of the disordered system. Its Fourier transform is $G_{QJ,Q'J'}(\omega) = \int dt G_{QJ,Q'J'}(t) e^{i (\omega+i \eta) t}$:
\begin{equation}
G_{QJ,Q'J'}(\omega) = \frac{\delta_{Q Q'} \delta_{J J'}}{(\omega + i \eta)^2 - \omega_{QJ}^2},
\end{equation}
$\eta$ being a positive infinitesimal value.

The decay of the Bloch mode $qj$ in the disordered system is described by the auto-correlation function of the Bloch displacements:
\begin{equation}
\tilde{G}_j(q,t) = \frac{\Theta(t)}{i \hbar}\Braket{0 | \left[ \hat{U}^0_{qj}(t), (\hat{U}^0_{qj})^\dagger \right] |0}.
\end{equation}
From here, the generalized unfolding formula for the spectral function is
\begin{equation}
\tilde{A}_j(q,\omega) = -\frac{2 \omega}{\pi} \text{Im} \int dt e^{i (\omega+i \eta) t} \tilde{G}_j(q,t),
\end{equation}
which yields for $\omega>0$:
\begin{align}
\label{specfun_final}
& \tilde{A}_j(q,\omega) = \\
& \sum_J  \left| \sum_{\alpha i r} \frac{e^{-i q r}}{\sqrt{N}} (e^\alpha_{i,qj})^* \sqrt{\frac{m^0_i}{m_{ir}}} V^\alpha_{ir,Q(q)J} \right|^2 \delta(\omega - \omega_{Q(q)J}). \nonumber
\end{align}

Note that if the factor $\frac{m^0_i}{m_{ir}}$ is omitted in equation~\eqref{specfun_final} in the presence of mass disorder, then what is computed is the spectrum of the disordered vibration mode 
\begin{align}
\hat{W}^0_{qj} & =  \frac{1}{\sqrt{N_\text{sc} N}} \sum_{i r \alpha} e^{-i q r}  (e^\alpha_{i,qj})^* \; \hat{U}^{\alpha}_{ir} \nonumber \\ & = \frac{1}{\sqrt{N_\text{sc} N}} \sum_{i r \alpha} e^{-i q r}  (e^\alpha_{i,qj})^* \; \sqrt{m_{ir}}\hat{u}^{\alpha}_{ir}.
\end{align}
The spectral function is then defined as 
\begin{align}
& A_j(q,\omega) = \\
& -\frac{2 \omega}{\pi} \text{Im} \int dt e^{i (\omega+i \eta) t} \frac{\Theta(t)}{i \hbar}\Braket{0 | \left[ \hat{W}^0_{qj}(t), (\hat{W}^0_{qj})^\dagger \right] |0}, \nonumber
\end{align}
yielding for $\omega>0$:
\begin{equation}
\label{specfun_final_wrong}
A_j(q,\omega) = \sum_J  \left| \sum_{\alpha i r} \frac{e^{-i q r}}{\sqrt{N}} (e^\alpha_{i,qj})^* V^\alpha_{ir,Q(q)J} \right|^2 \delta(\omega - \omega_{Q(q)J}).
\end{equation}

For the simple models considered in our study, equations~\eqref{specfun_final_wrong} and ~\eqref{specfun_final} reduce to equations~(2) and~(3) of the manuscript, respectively.

\section{The Green-Kubo formula and the spectral function}
\label{appendix_kubo}

In the absence of many-body interactions between phonons, the lattice thermal conductivity is given by the Green-Kubo formula. It can be expressed as (see eq. (2.50), (2.79b) and (2.83) in Ref.~\onlinecite{elliottTheoryPropertiesRandomly1974}, note the different definition for the Green's function):
\begin{equation}
\bm{\kappa} = \int_0^\infty d\hbar \omega \frac{4 \hbar \omega^2}{\pi V T} \left( -\frac{\partial f_B}{\partial\hbar\omega} \right) \text{Tr} \left[ \text{Im}(G) \bm{S} \; \text{Im}(G) \bm{S} \right]
\end{equation}
where $V$ is the system volume, $f_B$ is the Bose-Einstein distribution, $G (\omega) = \frac{1}{(\omega + i \eta)^2 - D}$ is the Green's function matrix, and $\bm{S}^{\alpha {\alpha'}}_{ir,i'r'} = \frac{1}{2i}(\bm{r}+\bm{r}_i-\bm{r'}-\bm{r}_{i'}) D^{\alpha {\alpha'}}_{ir,i'r'}$ is the heat current matrix, with $\bm{r}_i$ the position of atom $i$ inside the central unit cell. If we neglect the force disorder in $\bm{S}^{\alpha {\alpha'}}_{ir,i'r'}$ or if only mass disorder is present, then $\bm{S}^{\alpha {\alpha'}}_{ir,i'r'} = \sqrt{\frac{m^0_{i}}{m_{ir}}} \bm{S}^{0 \, \alpha {\alpha'}}_{ir,i'r'}\sqrt{\frac{m^0_{i'}}{m_{i'r'}}}$ with $\bm{S}^{0 \, \alpha {\alpha'}}_{ir,i'r'}$ the heat current matrix in the clean system.  
The Green's function matrix is diagonal in the disordered eigenbasis:
\begin{align}
& G^{\alpha {\alpha'}}_{IR,I'R'} =   \\
& \sum_{QJ} V^\alpha_{I,QJ} \frac{e^{i Q R}}{\sqrt{N_\text{sc}}} \frac{1}{(\omega + i\eta)^2 - \omega_{QJ}^2} \frac{e^{-i Q R'}}{\sqrt{N_\text{sc}}} (V^{\alpha'}_{I',QJ})^* \nonumber
\end{align}
and the clean heat flux is diagonal in the Bloch basis (we neglect interband terms \cite{allenThermalConductivityDisordered1993}):
\begin{equation}
\bm{S}^{0 \, \alpha {\alpha'}}_{ir,i'r'} = \sum_{qj} E^\alpha_{ir,qj} \omega^0_{qj} \bm{v}^0_{qj} (E^{\alpha'}_{i'r',qj})^*
\end{equation} 
where $\omega^0_{qj}$ and $\bm{v}^0_{qj}$ are the frequency and the group velocity of the eingenmode $qj$ in the clean system. 
Thus, taking the trace on the Bloch vector basis $E_{qj}$:
\begin{align}
\bm{\kappa} = & \int_0^\infty d\hbar \omega \frac{4 \hbar \omega^2}{\pi V T} \left( -\frac{\partial f_B}{\partial\hbar\omega} \right) \\
 & \times \sum_{qj q'j'}  \left[ E^\dagger_{qj} \sqrt{\frac{M^0}{M}} \text{Im}(G) \sqrt{\frac{M^0}{M}} E_{q'j'} \right] \nonumber \\ 
& \times \left[  E^\dagger_{q'j'}  \sqrt{\frac{M^0}{M}} \text{Im}(G) \sqrt{\frac{M^0}{M}} E_{qj} \right] \omega^0_{qj} \bm{v}^0_{qj} \omega^0_{q'j'} \bm{v}^0_{q'j'} \nonumber
\end{align}
where $M_{IR,I'R'} = \delta_{RR'} \delta_{II'} m_I$ is the matrix of the disordered masses and $M^0_{ir,i'r'} = \delta_{rr'} \delta_{ii'} m^0_i$ is the matrix of the clean masses.
If we neglect the cross-terms $q' \neq q$ (i.e. we neglect the interferences between the two Green's functions, that is to say the vertex corrections), then the thermal conductivity along the normalized vector $\bm{u}$ becomes:
\begin{align}
\kappa = & \sum_q \frac{\hbar (\omega^0_q v^0_q)^2}{\pi V T} \int_0^\infty d\hbar \omega  \left( -\frac{\partial f_B}{\partial\hbar\omega} \right) \nonumber \\
& \times \Bigg[  \sum_J  \left| \sum_{\alpha i r} \frac{e^{-i q r}}{\sqrt{N}} (e^\alpha_{i,qj})^* \sqrt{\frac{m^0_i}{m_{ir}}} V^\alpha_{ir,Q(q)J} \right|^2 \nonumber \\ 
& \times \text{Im} \left( \frac{2 \omega}{(\omega + i\eta)^2 - \omega_{QJ}^2} \right) \Bigg]^2 
\end{align}
where $v^0_q = \bm{v}^0_q \cdot \bm{u}$. Noting that $ \text{Im} \left( \frac{2 \omega}{(\omega + i\eta)^2 - \omega_{QJ}^2} \right) = - \pi \delta(\omega - \omega_{QJ})$, we find:
\begin{equation}
\kappa = \sum_q \frac{\pi \hbar (\omega^0_q v^0_q)^2}{V T} \int_0^\infty d\hbar \omega  \left( -\frac{\partial f_B}{\partial\hbar\omega} \right) \tilde{A}_j(q,\omega)^2.
\end{equation}

\section{The Chebyshev polynomial Green's function method}
\label{appendix_CPGF}

The Chebyshev polynomials Green's function (CPGF) approach has been reviewed in Refs.~\onlinecite{ferreiraCriticalDelocalizationChiral2015,weisseKernelPolynomialMethod2006} for the case of electrons, and has been adapted in Ref.~\onlinecite{bouzerarDrasticEffectsVacancies2020} to investigate phonon lifetimes in defected graphene. Here a brief overview of the method will be given. The phonon Green's function of a large disordered supercell is expanded on the Chebyshev polynomial basis:
\begin{equation}
\label{CPGF_expansion}
G(\bar{\omega}) = \sum^\infty_{n=0} g_n((\bar{\omega} + i \bar{\eta})^2) T_n(\bar{D})
\end{equation}
where the bar indicates that the spectrum has been rescaled to $[-1,1]$, the $g_n(z)$ are known complex functions:
\begin{equation}
g_n(z) = -i (2-\delta_{n,0}) \frac{(z - i\sqrt{1-z^2})^n}{\sqrt{1-z^2}}
\end{equation}
and the $T_n(\bar{D})$ are the Chebyshev polynomials evaluated at the dynamical matrix, that follow the recursion relation $T_{n+1}(\bar{D}) = 2 \bar{D} T_{n}(\bar{D}) - T_{n-1}(\bar{D})$. 

The spectral function~\eqref{specfun_final} for the Bloch mode $qj$ can be expressed as 
\begin{equation}
\tilde{A}_j(q,\bar{\omega}) = -\frac{2 \bar{\omega}}{\pi}\text{Im}\left( E^\dagger_{qj} \sqrt{\frac{M^0}{M}} G(\bar{\omega}) \sqrt{\frac{M^0}{M}} E_{qj} \right),
\end{equation}
so in practice, the quantities to be calculated are the so-called moments $\mu_{n,qj}$:
\begin{equation}
\mu_{n,qj} = E^\dagger_{qj} \sqrt{\frac{M^0}{M}} T_n(\bar{D}) \sqrt{\frac{M^0}{M}} E_{qj},
\end{equation}
which are computed using the recursion relation between the $T_{n}(\bar{D})$. 

The number of moments necessary for the sum~\eqref{CPGF_expansion} to converge is roughly equal to $1/2\bar{\omega}\bar{\eta}$. Because $\bar{\eta}$ is an artifical broadening and should be smaller than the disorder-induced spectral linewidth, probing modes closer and closer to $\Gamma$ requires more and more polynomials to be included.

\section{First principles thermal conductivity calculation on SiGe and MoS$_2$ compounds.}
\label{appendix_SiGe_MoS2}

\begin{figure}
\centering
\includegraphics[width=0.9\columnwidth]{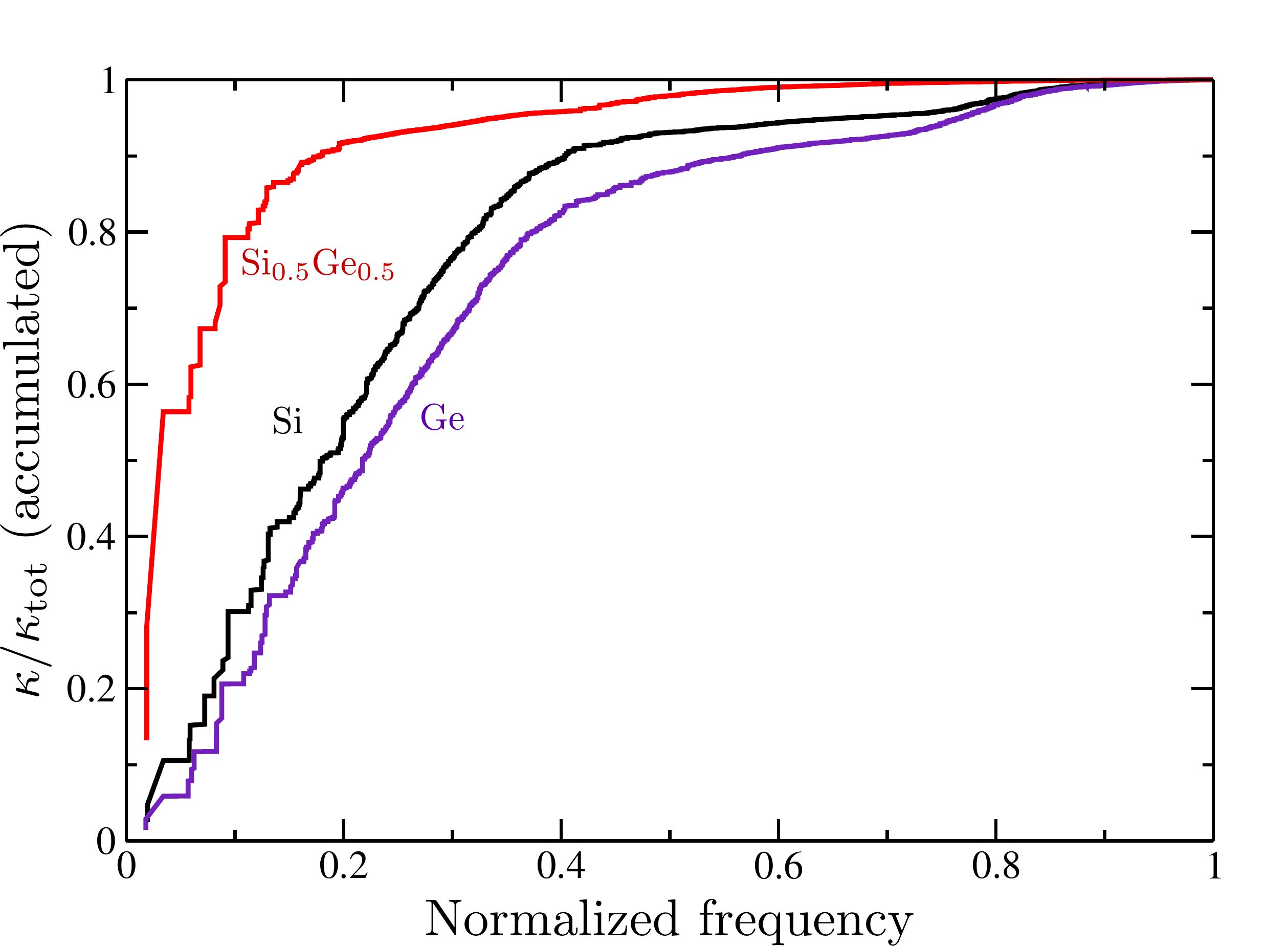}
\caption{Calculated room temperature normalized thermal conductivity accumulation as a function of frequency for Si (black curve), Ge (purple curve), and a Si$_{0.5}$Ge$_{0.5}$ alloy (red curve). The thermal conductivities are scaled to the corresponding calculated room temperature bulk values (including phonon-isotope scattering): \SI{144.9}{W/m.K} for Si, \SI{60.7}{W/m.K} for Ge, and \SI{9.7}{W/m.K} for Si$_{0.5}$Ge$_{0.5}$. The frequencies are scaled by the highest calculated frequency in each system:  \SI{15.6}{THz} for Si, \SI{9.0}{THz} for Ge, and \SI{11.2}{THz} for Si$_{0.5}$Ge$_{0.5}$ within the virtual crystal approximation.}
\label{figSiGe}
\end{figure}

Fig.~\ref{figSiGe} gives the calculated thermal conductivity accumulation as a function of phonon frequency:
\begin{equation}
\label{boltzmann}
\kappa (\omega) = \sum_{qj} C_{qj} v_{qj}^2 \tau_{qj} \Theta(\omega - \omega_{qj})
\end{equation}
for Si, Ge, and a Si$_{0.5}$Ge$_{0.5}$. Here, $\omega_{qj}$ is the frequency of phonon mode with wavevector $q$ and polarization $j$, $C_{qj}$ is the mode specific heat, $v_{qj}$ is the phonon velocity in the direction of an applied heat flux, $\tau_{qj}$ is the phonon transport time in this direction, and $\Theta$ is the Heaviside step function. The conductivities and frequencies are normalized as described in the figure caption. For Ge and Si, the harmonic and anharmonic interatomic force constants that determine the phonon properties in Eq.~\eqref{boltzmann} are determined from density functional theory as described in Ref.~\onlinecite{lindsayInitioThermalTransport2013}. For these calculations, full solution of the Boltzmann transport equation \cite{lindsayInitioThermalTransport2013,peierlsQuantumTheorySolids2001} determines the phonon lifetimes as limited by three-phonon and phonon-isotope interactions \cite{tamuraIsotopeScatteringDispersive1983}. For Si$_{0.5}$Ge$_{0.5}$ all properties (harmonic and anharmonic force constants, masses, lattice constants, etc.) of separate Si and Ge systems were averaged as in the virtual crystal approximation as described in Ref.~\onlinecite{liThermalConductivityBulk2012}. Additional phonon scattering from Si/Ge mass disorder was included via perturbation theory exactly as done for phonon-isotope interactions \cite{tamuraIsotopeScatteringDispersive1983,liThermalConductivityBulk2012}.

\begin{figure}
\centering
\includegraphics[width=0.9\columnwidth]{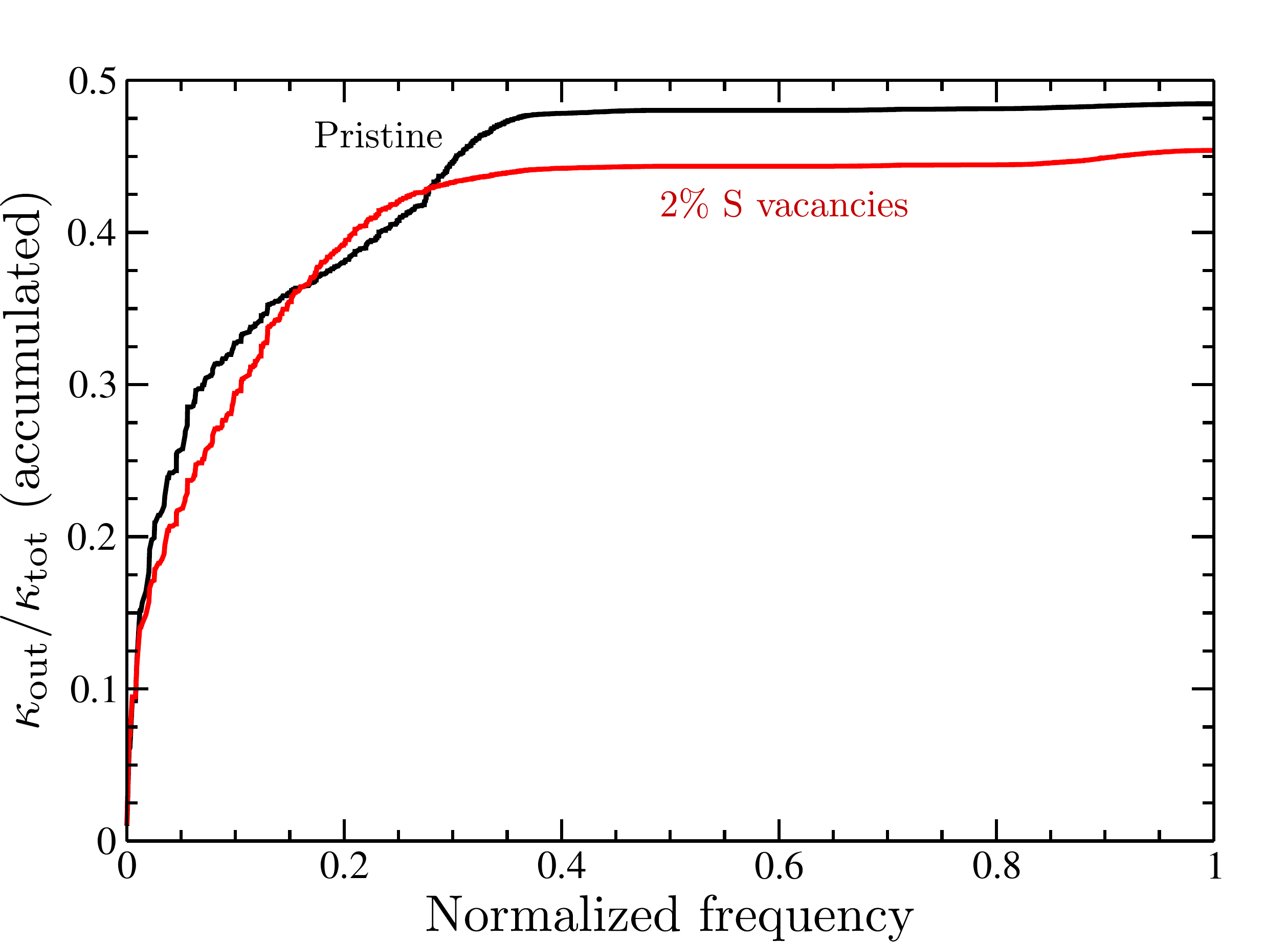}
\caption{Calculated out-of-plane contribution to the room temperature normalized thermal conductivity accumulation as a function of frequency for pristine monolayer MoS$_2$ (black curve) and monolayer MoS$_{1.96}$ (red curve). The thermal conductivities are scaled to the corresponding calculated room temperature bulk values: \SI{157.3}{W/m.K} for pristine MoS$_2$ and \SI{40.6}{W/m.K} for defected MoS$_2$. The frequencies are scaled by the highest calculated frequency in MoS$_2$: \SI{13.7}{THz}.}
\label{figMoS2}
\end{figure}

Fig.~\ref{figMoS2} gives the calculated contribution from out-of-plane phonons to the thermal conductivity accumulation as a function of phonon frequency for monolayer MoS$_2$ and monolayer MoS$_{1.96}$ (2\% S vacancies). The contribution from out-of-plane phonons is calculated by multiplying the contribution from each mode by the component of the polarization vector along the out-of-plane direction. Details of the first principles thermal conductivity calculations for MoS$_2$ are given in Ref.~\onlinecite{polancoDefectlimitedThermalConductivity2020}

\section{Other computational details}

For the CPGF calculations of the spectral function, we typically include \num{1.5e6} moments and \num{8e6} atoms in our supercells. For such system sizes, it is unnecessary to ensemble-average over multiple disorder configurations. The inverse phonon lifetimes are extracted by evaluating the full-width at half maximum (FWHM) of the quasiparticle peaks in the spectral function. This is done by fitting the peaks by the product of a Lorentzian and a linear function, then a spline interpolation is performed to easily calculate the FWHM.

The partial density of states is obtained via exact diagonalization of the dynamical matrix of 100 supercells with 400 atoms on average of which 200 are impurities. In addition to randomly distributing the impurities within the supercell, also the shapes of the supercells are randomized under the constraint that the number of atoms lie between 375 and 425 and the angles between the vectors that span the supercell are within 75 and 105 degrees.

%\begin{figure}[h]
%    \centering
%        \subfloat[Comparison of the DOS.]{\includegraphics[width=0.9\columnwidth]{figures/dos_abinitio_vs_modele} \label{sto_dope_V_comparison_model_dft_dos}}\\
%	\centering       
%        \subfloat[Comparison of the TDF assuming a constant mean free path of \SI{20}{\angstrom}.]{\includegraphics[width=0.9\columnwidth]{figures/TDF_abinitio_vs_modele_MFP20_eta0p0001} \label{sto_dope_V_comparison_model_dft_tdf}}
%    \caption{Comparison of the DOS and TDF of a $12.5 \%$ V-doped SrTiO$_3$ supercell between fully \textit{ab initio} calculations (thick blue line) and the model Hamiltonian (thin red line).}
%    \label{sto_dope_V_comparison_model_dft}
%\end{figure}

\begin{footnotesize}
\bibliographystyle{plain} 
\bibliography{main_manuscript.bbl}

\end{footnotesize}

\end{document}